\documentclass{article}
\title{Classical Electrodynamics with Vacuum Polarization}
\author{S. M. Blinder}
\date{ }

\begin{document}
\newcount\eq
\def\eqn{\eqno(\the\eq)}
\def\ad{\advance\eq by1}

\centerline{\bf Classical electrodynamics  with vacuum polarization:}
\centerline{\bf electron self-energy and radiation reaction} 
\bigskip
\centerline{S. M. BLINDER}
\centerline{University of Michigan, Ann Arbor, MI 48109-1055, USA}
\centerline{(e-mail: sblinder@umich.edu)}
\bigskip
 
\begin{abstract}
{\noindent The region very close to an electron  ($r\,
{\scriptscriptstyle{\buildrel <\over \sim}}\, r_0 = e^2/mc^2 \approx
2.8\times 10^{-13}$ cm) is,  according to quantum electrodynamics, a
seething maelstrom of virtual electron-positron pairs flashing in and out of
existence.  To take account of this well-established physical reality, a
phenomenological representation for vacuum polarization is introduced
into the framework of classical electrodynamics. Such a model enables a
consistent picture of classical point charges with finite electromagnetic
self-energy.  It is further conjectured that the reaction of a point charge to
its own electromagnetic field is tantamount to interaction with its vacuum
polarization charge or ``aura."  This leads to a modification of the
Lorentz-Dirac equation for the force on an accelerating
electron,  a new differential-difference equation which avoids the
pathologies of preacceleration and runaway solutions.}
\end{abstract}
\bigskip

\section{Introduction}

The singularities in fields and energies associated with point charges
in classical electrodynamics has been a pervasive flaw in what has been an
otherwise beautifully complete and consistent theory.  An immense
number of attempts to address this problem have been based, roughly
speaking,  on one of the following lines of argument: (1) Actual point
charges do {\it not} exist---{\it real} particles have a finite size---hence the
problem is artificial; (2) By a clever limiting procedure in the formalism,
the radius of a charge can be reduced to zero {\it without} introducing
infinities; (3) Point charges are quantum objects and classical
electrodynamics has no business dealing with them.
The last point of view, espoused by Frenkel\cite{1} and others, asserts that
any classical model is futile because the electron is a
quantum-mechanical object with no substructure. It is nonetheless of at
least academic interest to have a consistent classical relativistic model
which connects to macroscopic electrodynamics, while remaining
cognizant of its limitations.  The purpose of the present paper  is a
modified theory able to handle the singularities produced by point charges
while reducing to standard electrodynamics for $r\,\gg \,  r_0$.  

We propose to provide possible finishing
touches to Maxwell's electromagnetism  {\it without} making any
{\it ad hoc} modifications of the fundamental equations of the theory. The
key to our approach is the physical reality of vacuum polarization in the
submicroscopic vicinity of charged elementary particles. We begin
therefore with a review of the problem in the context of particle physics.

\section{Structure of the Electron} 

Since the discovery of the electron by J.
J. Thomson a century ago, the structure of ``the first elementary particle" 
has been the subject of extensive theoretical contemplation by some of the
leading figures of 20th Century physics\cite{2}. The earliest models
(Thomson, Poincar\'e, Lorentz, Abraham, Schott)\cite{3} pictured the
electron as a finite charged sphere, on the scale of the classical electron
radius $r_0 = e^2/mc^2
\approx 2.818\times 10^{-13}\, {\rm cm}$. The electromagnetic self
energy of such a finite structure would be of the order of
$W\approx {e^2/r_0}\approx mc^2 $  and thus implies an electron rest
mass predominantly electromagnetic in origin. 

Yet all experimental evidence implies an electron radius much smaller than
$r_0$, consistent, in fact, with a particle of point mass and point
charge\cite{4}.  Recent results of high-energy electron-positron
scattering\cite{5} imply an upper limit of $2\times 10^{-16}\,{\rm cm}$
on the electron size.

A number of ingeneous schemes to avoid a divergent electromagnetic
self-energy for a point electron have been proposed over the years by
Dirac\cite{6},  
Wheeler and Feynman\cite{7}, Rohrlich\cite{8}, 
Teitel\-boim\cite{9} and many others. The more recent approaches
invoke such arcana as {\it advanced} solutions of Max\-well's equations
(superposed on the conventional retarded solutions) and/or
renormalization of mass and charge infinities. This enables the divergent
part of the self-interaction to be avoided while leaving intact the radiation
reaction, an effect long known and thoroughly tested.

We will proceed on the premise that the electron rest mass (0.511
MeV/$c^2$) is {\it totally} electromagnetic, which was the original idea of
Lorentz and Abraham (see, however, Section 5). This is consistent with the
(nearly!) zero rest mass of the electron's uncharged weak isodoublet
partner---the neutrino---and with order of magnitude of the
neutron-proton mass difference (1.29 MeV/$c^2$). There is no need to
invoke any non-electromagnetic forces within the electron---collectively
known as Poincar\' e stresses. It should be noted that theories have been
proposed with counterbalancing {\it gravitational} fields\cite{10} but these
have been regarded with disfavor by Einstein\cite{11} among others.

\section {Stationary Point Charge}
 The energy of an electromagnetic field in a rest frame is given by
\ad$$W={1\over{8\pi}}\,\int\,\left({\bf E\cdot D + B\cdot
H}\right)\,d^3{\bf r}
\eqn$$\ad The field produced by a point charge $e$ in vacuum has
$D=E=e/r^2$ and
$$W= {1\over{8\pi}}\,\int\,{e^2\over{r^4}}\,4\pi r^2\, dr = \infty
\eqn$$\ad unless a lower cutoff is introduced.

It was suggested a long time ago by Furry and
Oppenheimer\cite{12} that quan\-tum-electrodynamic effects could give
the vacuum some characteristics of a polarizable
medium, which Weisskopf\cite{13} 
represented phenomenologically by an inhomogeneous dielectric constant,
viz
$$  {\bf D}(r) = \epsilon(r){\bf E}(r) \eqn$$\ad Accordingly,
$$W={1\over{8\pi}}\,\int_0^\infty\,{1\over\epsilon(r)}\,{e^2\over{r^4}}\,4\pi
r^2\, dr
\eqn$$\ad and equating this to the self-energy of the electron
$$W={e^2\over2}\,\int_0^\infty\,{dr\over{ r^2 \epsilon(r)}}=mc^2
\eqn$$\ad Remarkably, the functional form of $\epsilon(r)$ need not be
further specified, provided only that it satisfies the limiting conditions
$$\epsilon(\infty)=1 \qquad\hbox{and}\qquad \epsilon(0)=\infty
\eqn$$\ad 
Maxwell's first equation ${\bf\nabla\cdot E}=4\pi\varrho$ 
applied to the electric field 
$${\bf E}={e{\bf r}\over \epsilon(r) r^3}\eqn$$\ad 
determines the charge density
$$\rho(r)=-{e\, \epsilon'(r)\over {4\pi r^2[\epsilon(r)]^2} }\eqn$$\ad 
Note that this represents the {\it net} or {\it total} charge density, the sum of
the free and polarization densities. 
This function is appropriately normalized since
$$\int_0^\infty\,\rho(r)\, 4\pi
r^2\,dr=-e\,\int_0^\infty\,{\epsilon'(r)\,dr\over  {[\epsilon(r)]^2} } =
e\left[{1\over
\epsilon(\infty)}- {1\over \epsilon(0)}\right]=e \eqn$$\ad An explicit
functional form for $\epsilon(r)$ does follow if it is conjectured that the
net charge density (8) is proportional to the field energy
density from (5).  For then,
$${\epsilon'(r)\over  {\epsilon(r)} }=-{e^2\over2 mc^2 r^2}\eqn$$\ad  
with the
solution
$$\epsilon(r)=\exp\left({e^2\over 2mc^2r}\right) =\exp\left({r_0\over
2r}\right)
\eqn$$\ad

It should be emphasized for the benefit of QED theorists who might be
reading this that our use of the term ``vacuum polarization" is
intended only in a classical phenomenological context. The leading
contribution to vacuum polarization in real life comes from the interaction
of the electron with the transverse radiation field, which does not enter
in our model.  We are thereby overlooking
additional self-energy contributions arising from
fluctuations in the vacuum radiation field.  Accordingly, our representation
of vacuum polarization is {\it not} to be compared with QED computations.

Somewhat of a rationalization
for the functional form of $\epsilon(r)$ is suggested by Debye-H\"uckel theory
for ionic solutions and plasmas. The dielectric constant depends on a
Boltzmann factor
$e^{-{\cal E}/kT}$. If in place of the average thermal energy
$kT$, we  substitute the relativistic energy of pair formation $2mc^2$,
regarding the vacuum as an effective thermal reservior, then Eq (11) follows
with
${\cal E}=e^2/r$.

An explicit expression for the charge density follows by substituting (11)
into (8):
$$\rho(r)={er_0\over{8\pi r^4}}e^{-r_0/2r}  \eqn$$\ad Since
$\rho_{\rm free}(r)=e\delta({\bf r})$,  the density from vacuum
polarization must equal
$$\rho_{\scriptscriptstyle\rm VP}(r)={er_0\over{8\pi
r^4}}e^{-r_0/2r}-e\delta({\bf r}) \eqn$$\ad 
According to this model, the free point charge is
exactly cancelled by the deltafunction term of the polarization charge. The
corresponding electrostatic potential is given by
$$\Phi({\bf r})={2e\over r_0}\left(1-e^{-r_0/2r}\right)\,\approx\,
{e\over r} 
\qquad\hbox{when}\,\cases{   r_0\,\to\,0 \cr\quad\hbox{or}\cr
r\,\to\,\infty}     \eqn$$\ad This implies a deviation from
Coulomb's law of the same magnitude as the fine structure in atoms, but
totally negligible on a macroscopic scale.
An alternative evaluation of the electromagnetic self-energy follows
from  transformation of Eq (1) as follows:
$$W= {1\over{8\pi}}\,\int\,{\bf E\cdot D}\,d^3{\bf r}
={1\over 2}\,\int\,\Phi_{\rm free}\,\rho\, \,d^3{\bf r}
\eqn$$\ad using 
$${\bf D}=-\nabla\Phi_{\rm free}={e\,{\bf r}\over r^3}\eqn$$\ad   
and assuming
the requisite vanishing of integrands at infinity.  Thus
$$W={1\over 2}\,\int_0^\infty\,\Phi_{\rm free}(r)\rho(r)\,4\pi
r^2\,dr={e^2 r_0\over 4}\,\int_0^\infty\,{{e^{-r_0/2r}}\over r^3}\,dr
=mc^2
\eqn$$\ad in agreement with the previous result, and further justification
for the conjectured functional form of $\epsilon(r)$.

The preceding result suggests that the self-interaction of an electron is in
some sense equivalent to the interaction between a point charge and its {\it
net} polarization density---which we will denote as its ``aura" (in New Age
jargon, an energy field which emanates from a body).   In the following
section, we will utilize this picture to derive the radiation reaction for an
accelerated charge.

\section{Accelerating Point Charge}
\noindent The Lorentz-Dirac equation for the force on an
accelerating electron is given by\cite{6}
$$F_{\rm ext}^\lambda=ma^\lambda-{2 \,e^2\over 3\, c^3}\left(\dot
a^\lambda+{1\over c^2} a^2 v^\lambda\right)  \eqn$$\ad 
However, this equation
has fallen into disfavor in recent years because it admits pathological
solutions, including preacceleration and runaway behavior\cite{14}. Such
unphysical behavior is the result of taking the limit of the electron radius to
zero. It can be avoided by treating the electron as a finite charged sphere,
leading to a differential-difference equation without such
pathology\cite{15}.  Our picture of the electron as a point charge
interacting with its aura provides such an extended structure in a physically
natural way.  

The configuration of the aura surrounding an accelerating electron is most
likely quite complicated.  
At the very least, the aura is distorted from its original spherical
symmetry by Lorentz contraction. In addition, complicated
processes involving creation and relaxation of vacuum polarization in the
vicinity of the accelerating electron are certain to be occurring. We shall
assume a highly idealized model for the aura, treating it as a point charge
trailing the electron at a distance $R^{\textstyle *}$ with a proper-time
delay  $\tau^{\textstyle *}$.    Analogously, the simplest model for an ionic
crystal idealizes a lattice consisting of point charges.
We  will work in covariant notation throughout, thus avoiding the
``4/3 problem" and other relativistic pitfalls.  The
 Li\'enard-Wiechert 4-potential for a moving point charge is given by
 
$$A^\lambda(x)=e\,\left[{v^\lambda}\over{v\cdot R}\right]_{\rm
ret}
\eqn$$\ad  and the corresponding field tensor is 
$$\displaylines{F^{\lambda\mu}=\partial^\lambda A^\mu-\partial^\mu
A^\lambda=
\bigg[{e\over(v\cdot R)^2}\left(R^\lambda a^\mu-a^\lambda
R^\mu\right)
\hfill\cr\hfill +{e\over(v\cdot R)^3}\left(c^2-a\cdot
R\right)\left(R^\lambda v^\mu-v^\lambda R^\mu\right) \bigg]_{\rm
ret}\qquad (20) \cr}$$\ad 
 Four-dimensional scalar products are expressed
$a\cdot b=a^\mu b_\mu = a_0b_0-{\bf a\cdot b}\,$  (``West Coast
metric"). The relevant variables are
$$R^\lambda=\,(R,{\bf R}),  \qquad\qquad  v^\lambda=(\gamma
c,\gamma{\bf v}), $$
$$a^\lambda={{d v^\lambda}\over{d\tau}}=\left(\gamma^4\,{\bf a\cdot
v}/c, \gamma^2\,{\bf a}+\gamma^4\,({\bf
a\cdot v})\,{\bf v}/c^2\right)\eqn$$\ad 
$\bf R$ is the
displacement from the charge at the retarded time to the field point at the
present time. Thus
$R^\lambda$, lying on the light cone, is a null 4-vector with
$R^\mu R_\mu=0$.  We will also require the relations

$$ \dot a^\lambda={{d
a^\lambda}\over{d\tau}} ,\qquad v^\mu v_\mu=c^2,  \qquad v^\mu
a_\mu=0\eqn$$\ad 

We picture the point charge representing the aura to be chasing the
electron along the same trajectory, with an effective time delay
$\tau^{\textstyle *}$ relative to the proper time $\tau$.  Additionally, 
let the displacement
$R^\lambda$  produced during the time $\tau^{\textstyle *}$ be
parametrized as
$$R^\lambda={[v^\lambda]_{\rm ret}\over c}\,R^{\textstyle *}\eqn$$\ad
in terms of  an effective separation $R^{\textstyle *}$ between the
electron and its aura (not necessarily to the center
of the aura).  The parameters
$\tau^{\textstyle *}$ and $R^{\textstyle *}$ are independent and
$R^\lambda$ is no longer restricted to the light cone since
$R^\lambda R_\lambda=R^{\textstyle *}{}^2\not\equiv 0$.

Substituting (23) into (20), noting that $v\cdot R=c\,R^{\textstyle *}$ and
writing $\tau-\tau^{\textstyle *}$  for the retarded time, we obtain a major
simplification to

$$F^{\lambda\mu}(\tau)={ e\over c^3 R^{\textstyle *}}\,[v^\lambda
a^\mu -v^\mu a^\lambda]_{\tau-\tau^* }\eqn$$\ad 
According to Lorentz and Abraham, if the electron is a
purely electromagnetic entity, the self force should exactly balance the
external force. Thus 
$$F^\lambda_{\rm ext}(\tau)=-F^\lambda_{\rm self}(\tau)=-{e\over
c}\,F^{\lambda\mu}(\tau)v_\mu(\tau)   \eqn$$\ad 
We obtain thereby a
differential-difference equation for the force on an accelerating electron:
$$F^\lambda_{\rm ext}(\tau)={ e^2\over c^4 R^{\textstyle
*}}\,[a^\lambda v^\mu -a^\mu
v^\lambda]_{\tau-\tau^*}\,v_\mu(\tau)\eqn$$\ad 

 The values of $R^{\textstyle *}$ and $\tau^{\textstyle *}$ can be inferred by
considering the nonrelativistic limit, as $\bf v$ and
$\tau^{\textstyle *}$ approach zero. Expanding  $[v^\lambda]$ and $
[a^\lambda]$ and doing the summations over
$\mu$, we obtain
$$F^\lambda_{\rm ext}\approx{e^2\over  c^2\,R^{\textstyle
*}}\,a^\lambda \,- \,{e^2\,\tau^{\textstyle *}\over 
c^4\,R^{\textstyle*}}\,\dot a^\lambda
\eqn$$\ad  
Since this should reduce to the original Abraham-Lorentz equation\cite{16}
$${\bf F}_{\rm ext}\approx m\,{\bf a}  - {2 \,e^2\over 3\,
c^3}\,{\bf \dot a}
\eqn$$\ad
(as well as Newton's second law when $\tau^{\textstyle *}=0$) we can
identify 

$$R^{\textstyle *}= { e^2\over mc^2}\equiv r_0,
\eqn$$\ad  
the classical electron radius, and
$$\tau^{\textstyle *}={2e^2\over 3mc^3}\equiv\tau_0
\eqn$$\ad 
Remarkably, the parameter
$\tau_0\approx 6.26\times 10^{-24}$ sec  is the same ``relaxation time"
that occurs in the integration of the Lorentz-Dirac equation---the
immeasurably brief time interval during which classical acausal behavior is
tolerated. 

Finally, writing $\beta^\lambda=v^\lambda/c$,   we obtain a compact
differential-difference formulation for the force on an accelerating electron:
$$F^\lambda_{\rm ext}(\tau)=m\, [a^\lambda \beta^\mu -a^\mu
\beta^\lambda]_{\tau-\tau_0}
\,\beta_\mu(\tau) \eqn$$\ad  
Expansion of the bracketed quantity for small $\tau_0$ reacquires the
conventional Lorentz-Dirac equation:
$$F^\lambda_{\rm ext}= ma^\lambda  -{2 \,e^2\over 3\, c^3}\left(\dot
a^\lambda+{1\over c^2} a^2 v^\lambda\right) +{\cal O}(\tau_0)
\eqn$$\ad noting that
$\dot a^\mu v_\mu= -a^\mu a_\mu=-a^2$.

The nonoccurrence of runaway solutions to the modified Lorentz-Dirac
equation (31) is easy to prove. In the absence of external forces,
$$[a^\lambda v^\mu -a^\mu v^\lambda]_{\tau-\tau_0}
\,v_\mu(\tau)=0\eqn$$\ad  
Premultiplying by $v_\lambda(\tau-\tau_0)$ and
summing, we obtain
$$a^\mu(\tau-\tau_0)v_\mu(\tau)=0\eqn$$\ad 
Writing out the components explicitly, using Eq (21),
 $$[\gamma^4\,{\bf a\cdot v}/c]_{\tau-\tau_0}\,\gamma c-
[\gamma^2\,{\bf a}+\gamma^4\,({\bf
a\cdot v})\,{\bf v}/c^2]_{\tau-\tau_0}\cdot\gamma{\bf
v}(\tau)=0\eqn$$\ad  Clearly, if ${\bf v}$ is not
identically zero for all $\tau$, then ${\bf a}=0$. But if
${\bf v}= 0$ for all $\tau$, then ${\bf a}=0$ again. Thus the only
solutions for zero external force have zero acceleration. Further, the
absence of preacceleration is strongly implied by the dependence on no
time variables other than
$\tau$ and $\tau-\tau_0$. By contrast, the Lorentz-Dirac
equation (18) contains the derivative of acceleration, which can be
approximated by the finite difference
$$\dot a^\lambda(\tau)\approx {1\over 2\tau_0}
\left[a^\lambda\right]_{\tau-\tau_0}^{\tau+\tau_0} \eqn$$\ad 
with the
possibility of preacceleration attributed to the occurrence of the time
variable
${\tau+\tau_0}$.   
Note that the occurrence or absence of preacceleration can
{\it not} be readily discerned from the expanded form (32) of the L-D
equation.

\section{Non-Electromagnetic Mass} Although we have emphasized the
case of a charged particle with  purely electromagnetic self energy, the
treatment can easily be generalized to include non-electromagnetic
contributions to mass. In place of $m$ in all preceding formulas, substitute
$m_{\scriptscriptstyle\rm EM}$. For example,
$r_0=e^2/m_{\scriptscriptstyle\rm EM}c^2 $. The total self-energy can
now be written
$$W=m_{\scriptscriptstyle\rm total}c^2=(m_{\scriptscriptstyle\rm bare}+
m_{\scriptscriptstyle\rm EM})c^2 \eqn$$\ad 
 This might pertain to
elementary charged particles such as the muon, tauon, quarks and W
bosons---and possibly even to the electron if one accepts, for example, the
QED computation\cite{17} giving
$$W_{{\scriptscriptstyle\rm QED}}\approx {3\alpha\over{2\pi}}\, m
c^2\,\log\left({M\over m}\right)\eqn$$\ad 
 where $M\gg m$, defines a
relevant mass scale.


\bigskip\bigskip
\begin{thebibliography}{99}

\bibitem{1}  J. Frenkel J  {\it Zeit. Phys.} {\bf 32}  518 (1925).

\bibitem{2} For a retrospective on the electron centennial, see
S. Weinberg ``The first elementary particle"  {\it Nature} {\bf
386} 213 (1997).
 
\bibitem{3} A definitive review of classical electron theories is given by
F. Rohrlich  {\it Classical Charged  Particles} (Addison-Wesley, Reading,
MA, 1990).

\bibitem{4} See, for example D. H. Perkins {\it Introduction to High
Energy Physics}  (Addison-Wesley, Reading, MA, 1987).

\bibitem{5}   I. Levine  {\it et al}  {\it  Phys Rev. Lett.} {\bf 78} 424 (1997);  
see the artist's conception of the electron's structure on the AIP
Physics News website:  {\bf
http://www.aip.org/physnews/graphics/html/bare.htm}.

\bibitem{6}  P. A. M. Dirac   {\it Proc.
Roy. Soc. (London)}  {\bf A167} 148 (1938).

\bibitem{7}  J. A. Wheeler  and R. P. Feynman R P 
{\it Revs. Mod. Phys.} {\bf 17}
157; 1949  {\bf 21} 425 (1945).

\bibitem{8} F. Rohrlich   {\it  Phys. Rev. Lett. } {\bf 12} 375 (1964).

\bibitem{9}  C. Teitelboim  {\it  Phys. Rev. D} {\bf 1} 1572 (1970); 
  {\bf 3} 297 (1970 );   {\bf 4} 345 (1971 ).

\bibitem{10}  R. Arnowitt,  S. Deser and C. W. Misner {\it Phys. Rev.
Lett.} {\bf 4} 375 (1960 );  {\it Phys. Rev.} {\bf 120} 313 (1960).

\bibitem{11}  A. Einstein  {\it Sitzungsberichte Preussiche Akademie
der Wissenschaften Physikalische Mathematische Klasse} (1919); translation
in A. Einstein, {\it et al}  {\it The Principle of Relativity} (Dover, New York,
1952)

\bibitem{12}  W. Furry and  J. R. Oppenheimer  {\it Phys. Rev.} {\bf
45} 245 (1934).

\bibitem{13} V. F. Weisskopf   {\it Det. Kgl. Danske
Videnskab. Selskab. Mat.-Fys. Medd.}  {\bf 14} 1 (1936);  Reprinted in 
J. Schwinger  {\it Quantum Electrodynamics}  (Dover, New York, 1958).

\bibitem{14} J. L. Jim\'enez and I. Campos {\it Am. J. Phys.} {\bf 55} (1987).

\bibitem{15}  A. D. Yaghian {\it Relativistic Dynamics of a Charged
Sphere} (Springer, Berlin, 1992);  P. Caldirola   {\it Nuovo Cimento} {\bf
3}, Suppl. 2, 297 (1956); E. J. Moniz and  D. H. Sharp  {\it Phys. Rev.}
{\bf 15} 2850 (1977); F. Rohrlich {\it Am. J. Phys.} {\bf 65} 1051 (1997).

\bibitem{16}  For details of the derivation, see J. D. Jackson {\it Classical
Electrodynamics, 3rd Ed.} (Wiley, New York, 1999) Chap. 16. 

\bibitem{17} V. F. Weisskopf  {\it  Phys. Rev.} {\bf 56} 72 (1939).
\end{thebibliography}
\end{document}